\documentclass[conference]{IEEEtran}
\usepackage{amsmath}
\pdfoutput=1
\usepackage[pdftex]{graphicx}
\DeclareGraphicsExtensions{.pdf}
\usepackage{url}
\usepackage{booktabs}
\usepackage{multirow}

\title{Acoustic Echo Cancellation using Residual U-Nets}
\author{\IEEEauthorblockN{J.~Silva-Rodr\'{i}guez$^{(1)}$, M.F.~Dolz$^{(2)}$, M.~Ferrer$^{(3)}$, A.~Castell\'{o}$^{(2)}$, V.~Naranjo$^{(4)}$,  G.~Pi\~{n}ero$^{(3)}$}
\IEEEauthorblockA{$^{(1)}$ Inst. Transport and Territory, Universitat Polit\`{e}cnica de Val\`{e}ncia, jjsilva@upv.es \\
      $^{(2)}$ Dept. Ingeniería y Ciencia de los Computadores, Universitat Jaume I, dolzm@uji.es \\
      $^{(3)}$ ITEAM, Universitat Polit\`{e}cnica de Val\`{e}ncia, gpinyero@iteam.upv.es\\
      $^{(4)}$ Inst. Invest. Innov. Bioingeniería, Universitat Polit\`{e}cnica de Val\`{e}ncia, vnaranjo@dcom.upv.es}}

\begin{document}
%
\maketitle
\begin{abstract}
This paper presents an acoustic echo canceler based on a U-Net convolutional neural network for single-talk and double-talk scenarios. U-Net networks have previously been used in the audio processing area for source separation problems because of their ability to reproduce the finest details of audio signals, but to our knowledge, this is the first time they have been used for acoustic echo cancellation (AEC). The U-Net hyperparameters have been optimized to obtain the best AEC performance, but using a reduced number of parameters to meet a latency restriction of $40$\,ms. The training and testing of our model have been carried out within the framework of the \textit{ICASSP 2021 AEC Challenge} organized by Microsoft. We have trained the optimized U-Net model with a synthetic dataset only (S-U-Net) and with a synthetic dataset and the single-talk set of a real dataset (SR-U-Net), both datasets were released for the challenge. The S-U-Net model presented better results for double-talk scenarios, thus their inferred near-end signals from the blind testset were submitted to the challenge. Our canceler ranked 12th among 17 teams,  and 5th among 10 academia teams, obtaining an overall mean opinion score of $3.57$.
\end{abstract}
%
%
\section{Introduction}
\label{sec:intro}

The video conferencing global market size was valued at \$3.85 billion USD in 2019 and is predicted to significantly grow during 2020 due to the COVID-19 pandemic~\cite{UCToday}. From the point of view of the customers, one of the most annoying experience is to hear their own echo from the other side of the line due to a malfunctioning of the acoustic echo cancellation (AEC) system. 

The basic theory of AEC systems was first developed by AT\&T Bell Labs~\cite{Son67}, and it was one of the earliest applications of adaptive filters \cite{Haykin2002}. The aim is to suppress the undesired echo produced by the acoustic coupling between a loudspeaker and a microphone, usually driven by the same device. However, advanced AEC systems are also capable of reducing background noise and enhancing the near-end speech captured by the microphone~\cite{Hansler2006}. Traditionally, echo cancellation is accomplished by adaptively identifying the echo path impulse response between the loudspeaker and the microphone and subtracting the estimated echo signal from the microphone signal. In practice, the canceler must also deal with non-stationary scenarios and non-linear effects over the signals. To combat these impairments, residual echo suppressors (RES)~\cite{Gustafsson1998,Kuech2007} and voice activity detectors for the far-end and near-end signals s~\cite{Jongseo99,Bene2000,iqbal2006} are usually included in the AEC system design.

Recently, AEC solutions have been proposed from the machine learning (ML) perspective. On the one hand, there are works that use neural networks in combination with classical adaptive cancelers in order to detect double-talk situations~\cite{iqbal2006}, to model the spectrum of the residual echo~\cite{Schwarz2013}, or to suppress it~\cite{Carbajal2018,Chen2020}. On the other hand, complete AEC solutions implemented by means of deep learning models can be found in the recent literature.  In~\cite{Zhang2018} the authors approach the echo suppression as a source separation problem, proposing a recurrent neural network with bidirectional long short-term memory (BLSTM) to estimate the best ratio mask in double-talk scenarios. Same authors propose in~\cite{Zhang2019} to add a convolutional recurrent network (CRN) to estimate the near-end speech, and in parallel they use a LSTM to detect its presence. In~\cite{Fazel2019}, deep gated recurrent neural networks are proposed  to estimate the near-end signal and the echo using multitask learning, and in~\cite{Fazel2020} a hybrid AEC combining a classical adaptive filter and a contextual attention module (CAM) is used. 

There is no definite AEC solution and the problem is still open. For this reason, Microsoft has organized the \textit{ICASSP 2021 Acoustic Echo Cancellation Challenge} ``to stimulate research in the area of AEC''~\cite{Sridar2020,AECweb}. We describe in this work an AEC system based on Residual U-Nets, which has achieved an overall subjective score of $3.57$ in the challenge final results~\cite{AECweb}. The U-Net architecture was firstly proposed in \cite{Ronneberger2015U-net:Segmentation} to solve image segmentation problems in medical applications, but they have also been used to separate the singing voice from the background sounds~\cite{Jansson2017, Kadandale2020}, given its capacity for recreating high-quality audio. However, to our knowledge, this is the first time that U-Nets are used for acoustic echo cancellation. The rest of the paper is as follows: we describe our proposed AEC system in Section~\ref{sec:model}, including an optimization of the U-Net main hyperparameters. A brief description of the databases released by Microsoft for the challenge~\cite{Sridar2020} and several experimental results are shown in Section~\ref{sec:experiments}, whereas Section~\ref{sec:conclusions} summarizes the main conclusions.


\section{Proposed model}
\label{sec:model}

The methodological core of the proposed approach is a U-Net convolutional neural network (CNN)~\cite{Ronneberger2015U-net:Segmentation} able to suppress the echo in different communication scenarios. In the context of the AEC, the microphone signal $y(n)$ is a mixture of the original audio $s(n)$, so-called near-end speech, and the far-end speech $x(n)$ such that:
\begin{equation}
\label{eq:aec}
y(n) =  s(n) + f(x(n)) + v(n) =  s(n) + d(n) + v(n) \, ,
\end{equation}
where $v(n)$ denotes background noise and $f(\cdot)$ is the function modeling the transformation of the far-end speech signal emitted by the loudspeaker and captured by the microphone, i.e., the echo $d(n)$. This model describes a double-talk scenario where $s(n)$ is present in $y(n)$, whereas a single-talk scenario is denoted when $s(n)=0$. It is usually assumed that $f(x(n))$ is a linear function such that $d(n) = h(n) \ast x(n)$, where $h(n)$ is the room impulse response (RIR) between the loudspeaker and the microphone, but we will also consider non-linear effects on $x(n)$, as clipping or impairments due to time-varying RIRs.


\subsection{Feature extraction}
\label{ssec:feat}

The U-Net model takes as input pre-processed segments of the far-end speech, $x(n)$, and the microphone signal, $y(n)$, and aims to apply a deep filtering function to infer the estimated near-end speech, $\hat{s}(n)$. The proposed AEC pipeline uses $160$-ms time frames of both $y(n)$ and $x(n)$. The time frame is formed by stacking $T_f=40$ ms of new speech samples, and $120$\,ms of previous samples to ensure the preservation of the temporal information. Then, their short-time Fourier transforms (STFT) are obtained using a Hanning window of size $W = 318$ samples and $75\%$ of overlap. Normalized magnitude spectrograms of size $160 \times 32$ from the near-end microphone signal, $Y$, and from the far-end speech, $X$, are then obtained and used as input features to the model, as it is shown in Fig.~\ref{f:fig2}. During the inference stage, the normalization process is inverted to obtain the magnitude of the predicted near-end speech spectrogram, $\hat{S}$. Then the preserved phase information of the $Y$ spectrogram is added to $\hat{S}$, and their inverse STFT is calculated. The newest period of $40$\,ms samples is then selected to build the estimated near-end speech $\hat{s}(n)$. In case of single-talk scenarios where $s(n)$ is not present, the output of the model would not have any content, although we denote their spectrogram as $\hat{S}$ for the sake of clarity.

\subsection{Residual U-Net}
\label{ssec:u-net}
\begin{figure}[t]
    \begin{center}
    \includegraphics[width=.48\textwidth]{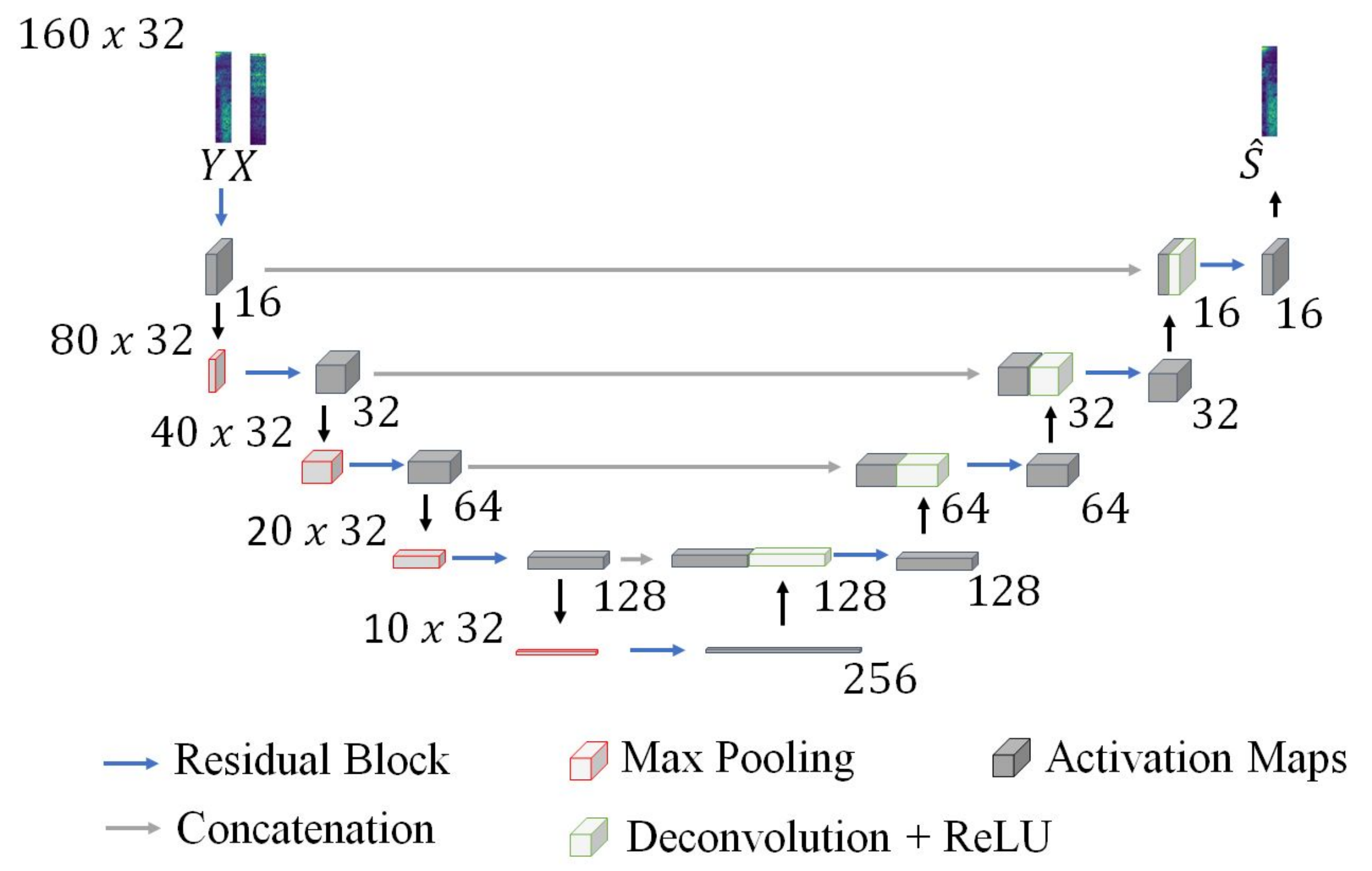}
    \caption{Residual U-Net proposed for acoustic echo cancellation.}
    \label{f:fig2}
    \end{center}
\end{figure}

As shown in Fig.~\ref{f:fig2}, the U-Net is composed of two branches: the encoder, responsible of obtaining relevant features from the image, and the decoder, in charge of the construction of the target image. The encoder branch consists of stacked convolutional blocks with increasing number of filters and dimensional reduction via the max-pooling operator. In the decoder branch, the convolutional blocks are followed by deconvolutions that progressively recover the spatial dimensions and reduce the number of filters to half. In order to combine multi-scale information, the encoder is connected via shortcut connections to the decoder. 
In our model, the spectrograms of the near-end microphone, $Y$, and the far-end speech, $X$, are used as inputs in a two-channel grayscale image of size $160 \times 32 \times 2$. The initial number of filters used, $F_0$, has been optimized as described in Section~\ref{ssec:tuning}. It is noteworthy to mention that the number of filters used is smaller than usual due to the strict rules of the AEC Challenge regarding real-time implementation~\cite{Sridar2020}. With respect to the pooling/deconvolution operations, the pooling is performed only in the frequency-related dimension of the spectrogams using filters of size $2 \times 1$. Due to the small size of the input time frames, a temporal pooling could soften the spectrogram and degrade the speech representation. As model output, a convolutional layer of size $1 \times 1$ reconstructs the estimated near-end speech spectrogram, $\hat{S}$. 

The convolutional block in charge of the feature extraction (blue connections in Fig.~\ref{f:fig2}) is designed using residual connections in the Residual U-Net. These connections have shown to improve deep learning models optimization, avoiding gradient vanishing problems in other U-Net applications \cite{roadExtraction,Kalapahar2020GleasonU-Net}. In our model, this block is composed of consecutive convolutional layers with size $3\times3$ and ReLU activation. The first layer standardizes the activation maps to the amount of filters corresponding to the depth level. The resultant activation maps from this layer are combined in a shortcut connection with the results of $N$-stacked convolutional layers via an adding operation. The number of stacked filters is empirically fixed during the hyperparameters tuning experiments described in Section~\ref{ssec:tuning}.

\subsection{Loss estimation}
\label{ssec:loss}
During the training stage, learning is driven by the comparison between the spectrograms of the predicted and the real (ground truth) near-end signals. Let us denote the spectrogram of the target signal as $S(m,k)$, where $m$ denotes the indices of the $M$ subframes of $W$ samples in $T_f$, and $k$ being the frequency bin, $k=0,1,\ldots,W/2$. The objective function is defined as the root-mean-square error between the spectrograms:
\begin{equation}
\label{eq:loss}
L = \sqrt{\frac{1}{M(W/2+1)}\sum_{m \in T_f,k}\left(\hat{S}(m,k) - S(m,k)\right)^2} \, .
\end{equation}

Thus, the network focuses on learning to filter the last $T_f$ time period of the near-end speech. Otherwise, the model could get stacked on a local minimum filtering the signal corresponding to the whole $160$\,ms frame.

\subsection{Tuning the U-Net model}
\label{ssec:tuning}
Given the large number of U-Net variants, we have performed a hyperparameter optimization search to select the model configuration that reduces both the model loss~\eqref{eq:loss} and the computational complexity of the inference stage, so as to accomplish the real-time requirements of the AEC Challenge. For this purpose, we have used Hyperas~\cite{pumperla2019hyperas}, a hyperparameter optimization tool for Keras models that operates on top of Hyperopt~\cite{bergstra2013hyperopt}. An advantage of Hyperas is that, instead of using a grid-based search, it leverages Bayesian algorithms to partially search the parameter space for relatively optimal parameter settings.
\begin{table}[t]
    \centering
    \caption{Values per each hyperparameter analyzed with Hyperopt for the U-Net model.}
    \label{tab:hyperas}
    \begin{tabular}{cc}
    \toprule
        Hyperparameter & Values \\ \midrule
        Optimizer & Nadam, SGD, Adam \\ 
        Learning rate & $10^{-3}$, $10^{-4}$, $10^{-5}$ \\ 
        \# Encoders-Decoders & 4--3, 3--2 \\ 
        Residual blocks & Conf\#1, Conf\#2 \\ 
        \# Filters ($F_0$) & 8, 16 \\ \bottomrule
    \end{tabular}
\end{table}

Table~\ref{tab:hyperas} displays the set of hyperparameters considered in the U-Net optimization. Specifically, we tested three different optimizers, namely SGD~\cite{bottou2012stochastic}, Nadam, and Adam~\cite{tato2018improving}, each with three different learning rate values, as shown in the second row of Table~\ref{tab:hyperas}. Besides, we considered the number of encoders and decoders of the U-Net (third row), and two different configurations for the residual blocks (fourth row): ``Conf\#1'' stacks two convolutional layers followed by the shortcut connection carrying the input, whereas ``Conf\#2'' follows the same principle as ``Conf\#1'', but adds a convolution operation in the shortcut prior to the addition of the residual connection to the main branch. The last hyperparameter (fifth row) considers two baseline values for the number of filters, $\{8,16\}$, scaled by a factor of $2^{i}$ and $2^{(n-i+1)}$ in the $i$-th encoder and $i$-th decoder of a U-Net, respectively. The result of this hyperparameter search revealed that the minimum loss was achieved with the Nadam optimizer using a learning rate of $10^{-4}$ on a U-Net consisting of 16 baseline filters, 4 encoders, 3 decoders and residual blocks according to ``Conf\#1''.

\section{Experiments and Results}
\label{sec:experiments}

\subsection{Real-time performance evaluation}
\label{ssec:RTperformance}

Once hyperparameters of the U-Net model were selected, we set up our target platform, a personal computer equipped with an Intel Core i5 quadcore running at 2.3\,GHz, to measure the inference time while using the TensorFlow Lite framework v2.0.1\footnote{https://www.tensorflow.org/lite}. As stated in the AEC Challenge rules~\cite{Sridar2020}, any proposed AEC system should take less than the stride time (in our case $40$\,ms) to process a frame, and the total algorithmic latency allowed should be $\leq 40$\,ms. Using 32-bits floating-point (FP32) arithmetic, the inference time was $45.63$\,ms, which exceeded the allowed latency. Then, we quantized the U-Net to use half-precision arithmetic (FP16). This allowed for a reduction of $52.5$\% with respect to using FP32. Table~\ref{tab:timebd} shows the time breakdown for the inference operation given an input frame when using FP16 (w/ quantization) and FP32 (w/o quantization).
 
\begin{table}[t]
    \centering
    \caption{Time breakdown of the complete inference operation.}
    \begin{tabular}{lcc}
    \toprule
         & Time w/ & Time w/o \\
        Stage & quantization (ms) & quantization (ms) \\ \midrule
        Get buffer ($40$\,ms) & $~~0.003$ & $~~0.003$ \\ 
        Data preparation & $~~1.193$ & $~~1,193$ \\ 
        Model inference & $18.953$ & $42.903$ \\ 
        Data extraction & $~~1.531$ & $~~1.531$ \\ \midrule
        Total & $21.68$ & $45.63$ \\ \bottomrule
    \end{tabular}
    \label{tab:timebd}
\end{table}


\subsection{Datasets}
\label{ssec:dataset}
Two training datasets were released by Microsoft for the AEC Challenge\footnote{For further details see~\cite{Sridar2020}. The AEC Challenge datasets are downloadable from: \url{https://github.com/microsoft/AEC-Challenge}}:

1) \textit{Real dataset}: This dataset consists of more than $2{,}500$ different real environments, audio devices, and human speakers in single-talk and double-talk scenarios with time-varying and stationary RIRs. Clean near-end speech segments were also provided, although they did not correspond to the near-end speech content present in the double-talk signals, i.e., there was no ground truth to feed the loss function~\eqref{eq:loss}.

2) \textit{Synthetic dataset}: It consists of $10{,}000$ synthetic samples representing single-talk, double-talk, near-end noise, far-end noise, and various nonlinear distortion situations. Each sample includes a far-end speech, echo signal, near-end microphone signal and the near-end speech that will be considered the ground truth $S$ in~\eqref{eq:loss}. Further details on each generated sample are available in the ``meta.csv'' file available in the repository regarding the  signal-to-echo-ratio (SER), the addition of noise to the near-end and far-end signal, and the non-linearity of the RIR. For the $10{,}000$ synthetic samples, the SER was uniformly distributed between $-10$ and $10$\,dB in steps of $1$\,dB.

Microsoft also released a \textit{Testset} at the beginning of the challenge consisting of approximately 800 recordings partitioned into the following subsets: ``Clean'', formed by clean speech with a Mean Opinion Score (MOS) $> 4$ for both near and far-end signals, and ``Noisy'', formed by noisy speech for both near and far-end signals. The signal conditions of this testset were similar to the blind testset used by Microsoft for the evaluation of the AEC Challenge, although the speakers and spoken texts were different. The estimated near-end signals inferred by the baseline model~\cite{Xia2020} were also released and we have used them as a reference for the performance comparison of the preliminary results presented in Section~\ref{ssec:pre_results}. 

\subsection{Performance metrics}
\label{ssec:metrics}
We have used the echo return loss enhancement (ERLE) and the perceptual evaluation of speech quality (PESQ)~\cite{ITU-P862} as the performance metrics. The ERLE evaluates the echo reduction achieved during single-talk periods, when the near-end signal $s(n)$ is not present:
\begin{equation}
  \mathrm{ERLE} = 10 \log_{10} \frac{E[y^2(n)]}{E[\hat{s}^2(n)]} \approx 10 \log_{10} \frac{\sum_{n} y^2(n)}{\sum_{n} \hat{s}^2(n)} \label{eq:erle}
\end{equation}
where the statistical expectation operation $E[\cdot]$ is estimated by the mean square value over the whole speech duration. The PESQ is an objective measure of the quality of the estimated near-end speech $\hat{s}(n)$ compared to that of the original speech $s(n)$, thus it will be used for double-talk scenarios. The PESQ scores range from $-0.5$ to $4.5$, and a higher score indicates a better speech quality. 

\subsection{Preliminary Results}
\label{ssec:pre_results}
We have trained the optimized U-Net model of Section~\ref{ssec:tuning} with the training datasets described above generating two different echo cancelers: 

\textit{S-U-Net}: Only the synthesized dataset has been used to train. A $10$\% of the dataset not used for training has been selected to be used as a testset in the performance analysis.   

\textit{SR-U-Net}: The synthesized dataset and the single-talk signals of the real dataset have been used to train the U-Net. The ground truth for the real dataset is a silence, that is, the pixels of the $S$ spectrogram all have a $0$ value.

We have also implemented a classic adaptive AEC filter of $4{,}000$ coefficients that uses a Partitioned Frequency-Block Least Mean Square (PFB-LMS) algorithm~\cite{Behrouz2013}. The block size was 1024 and the step size was $\mu=0.0001$. In case the algorithm starts to diverge, the step size is reduced by half. 

The ERLE values of the different methods are shown at the upper part of Table~\ref{tab:ERLE-PESQ}, where we have also included the results for the Testset baseline signals provided by Microsoft. For the synthetic testset, a mask from the near-end speech has been calculated to select the single-talk periods of the audio clips. In the case of the Testset (TS), only the signals labeled as single-talk have been used. It can be noticed that the ERLE levels shown by the SR-U-Net model on the Testset signals double the Baseline and S-U-Net levels, which is a result of the addition of the real single-talk signals to the training set. Regarding the PFB-LMS, it cannot reduce the echo more than a few dB in all cases.

\begin{figure}[t]
    \begin{center}
    \includegraphics[width=.48\textwidth]{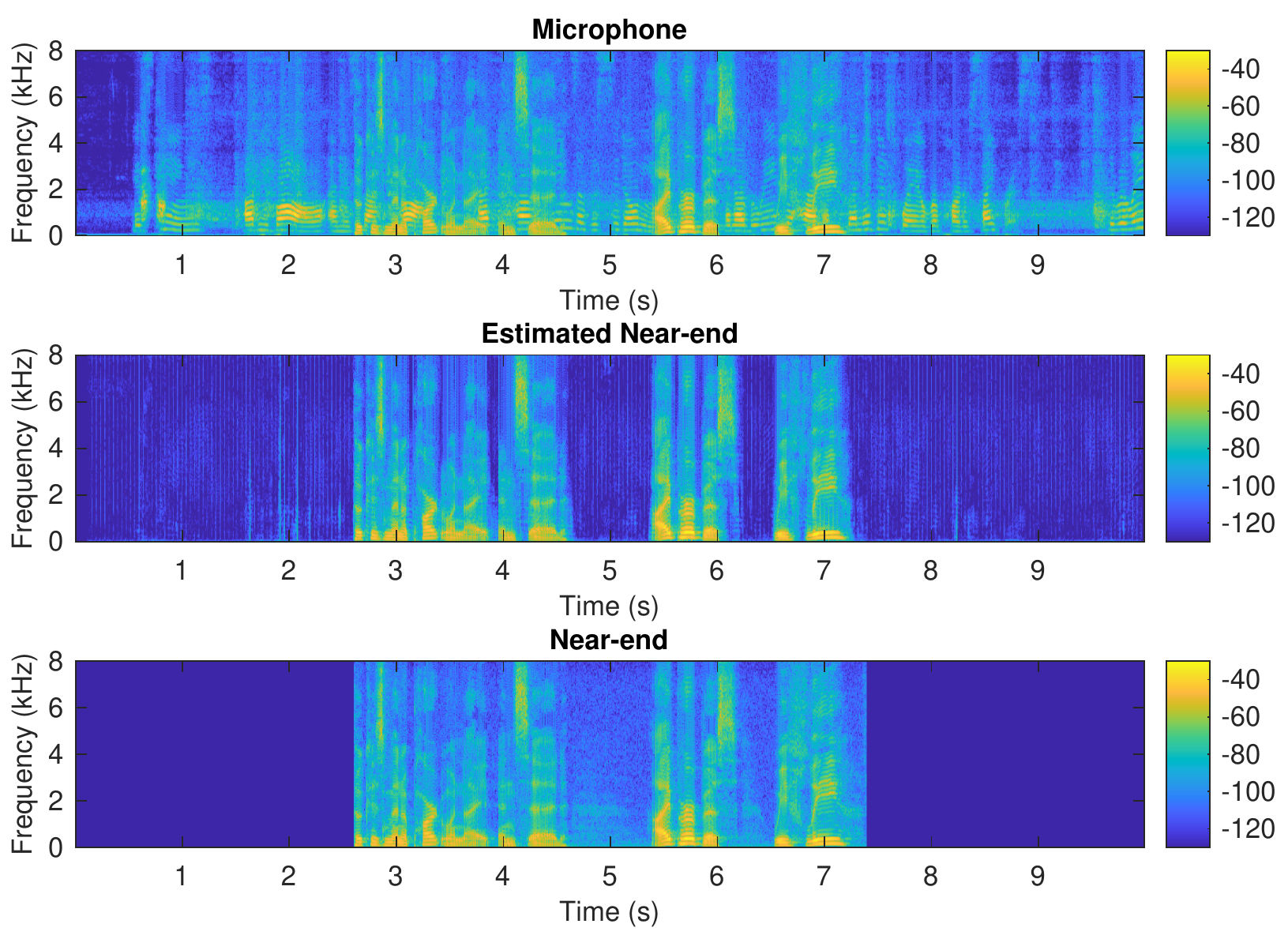}
    \caption{Spectrograms of microphone, estimated near-end, and near-end signals of synthesized sample \#$2477$.}
    \label{f:spectr}
    \end{center}
\end{figure}

Regarding the PESQ evaluation, it has been carried out for the synthetic testset taking the near-end signal (ground truth) as the reference. The unprocessed PESQ for the microphone signal has also been calculated as in~\cite{Fazel2019}, and the gain of each method with respect to the unprocessed PESQ is shown at the lower part of Table~\ref{tab:ERLE-PESQ}. A mask from the near-end speech has been calculated to select the double-talk periods of the audio clips. On the one hand, it can be appreciated how the extended training of the SR-U-Net model deteriorates its performance in double-talk scenarios. Therefore, the S-U-Net is preferred for its highest quality of the inferred near-end speech. Spectrograms in Fig.\ref{f:spectr} illustrate an AEC example on a signal from the synthetic testset. The microphone signal was formed using a nonlinear RIR, adding noise at the far-end, and adding the echo signal with a SER level is $1$\,dB. The near-end speech estimated by the S-U-Net achieved an ERLE of $39.9$\,dB and a PESQ score of $3.07$.
On the other hand, the PFB-LMS method achieves a PESQ gain similar to the S-U-Net, which does not correspond to the perceived quality when listening to their resulting inferred signals. It can be explained because the PESQ method~\cite{ITU-P862} was proposed to measure the distortions produced by a voice-over-IP network on a coded speech signal, but it is still in discussion its validity to measure the subjective perception when several speech signals are present~\cite{naderi2020open}.

\begin{table}[t]
    \centering
    \caption{ERLE values and PESQ gains of the AEC methods.}
    \label{tab:ERLE-PESQ}
    \begin{tabular}{ccccc}
    \toprule
         & Method & Synthetic & ``Clean'' TS & ``Noisy'' TS \\ \midrule
       & Baseline & X & $23.69$ & $25.00$ \\ 
  ERLE     & PFB-LMS & $~~6.27$ & $~~5.89$ & $~~4.20$ \\ 
    (dB)   & S-U-Net & $25.47$ & $19.29$ & $16.24$ \\ 
       & SR-U-Net & $32.71$ & $42.53$ & $44.31$ \\ 
 \bottomrule
    \end{tabular}
\begin{tabular}{cccc}  
 & & & \\
  \multicolumn{4}{c}{PESQ gain for the synthetic testset} \\ \midrule
        $0.00$ & $0.23$ & $0.28$ & $0.03$ \\
         (Unprocessed) & (PFB-LMS) & (S-U-Net) & (SR-U-Net) \\ \bottomrule
    \end{tabular}
\end{table}

Therefore, we submitted to the \textit{ICASSP2021 AEC Challenge} the near-end speech signals inferred by the optimized U-Net architecture obtained in Section~\ref{ssec:tuning} trained only with the synthetic dataset, i.e., the S-U-Net.

\subsection{AEC Challenge Results}
\label{ssec:chal_results}
Table~\ref{tab:CResults} presents the overall subjective scores obtained by our model in the AEC Challenge~\cite{AECweb}, together with the particular scores achieved for the ``Clean'' and ``Noisy'' subsets. The blind testset covered four different real scenarios: \textit{Single-talk Near-end (ST NE)} where only the near-end signal is present, \textit{Single-talk Far-end (ST FE) Echo} where only the echo due to the far-end signal is present, \textit{Double-Talk (DT) Echo} where the echo due to the far-end signal and the near-end signal are present, and  \textit{Double-Talk (DT) Others} where further impairments have been added to the \textit{DT Echo} signal conditions. The subjective score for the ST NE scenario has been the MOS obtained applying the ITU-T Recommendation P.808~\cite{naderi2020open}. The subjective score to evaluate the other three scenarios has been the Degradation Mean Opinion Score (DMOS) obtained by the application of the ITU-T Recommendation P.831~\cite{ITU-P831}. The overall score for each column is calculated as their mean value. Bold numbers indicate a score above the percentile 30, that is, $30$\% of the teams present a lower score. We ranked 12th among 17 teams, just below the Microsoft baseline. Among the teams from Academia, we ranked the fifth of ten Universities. From Table~\ref{tab:CResults}, it can be appreciated that our model presents a good behaviour in single-talk and double-talk scenarios, but it lacks of robustness for non-trained scenarios as the ``ST NE'' and ``DT Other'' sets. Given the limited synthetic dataset used for training, we are confident that the U-Net performance could be improved by expanding the training dataset. 
\begin{table}[t]
    \centering
    \caption{MOS achieved by our model in the AEC Challenge.}
    \label{tab:CResults}
    \begin{tabular}{cccc}
    \toprule
         Scenario & Clean & Noisy & Overall \\ \midrule
        ST NE (MOS) & $\mathbf{3.66}$ & $3.36$ & $3.51$ \\ 
        ST FE Echo (DMOS) & $\mathbf{4.05}$ & $\mathbf{3.52}$ & $\mathbf{3.79}$ \\ 
        DT Echo (DMOS) & $\mathbf{4.08}$ & $\mathbf{3.81}$ & $\mathbf{3.94}$ \\ 
        DT Other (DMOS) & $3.09$ & $3.02$ & $3.06$ \\ \midrule
        Overall & $\mathbf{3.72}$ & $\mathbf{3.43}$ & $\mathbf{3.57}$ \\ \bottomrule
    \end{tabular}
\end{table}

\section{Conclusion}
\label{sec:conclusions}

We presented in this work the AEC system used to obtain the estimated near-end speech signals submitted to the \textit{ICASSP 2021 AEC Challenge} as the ``Universitat Polit\`{e}cnica de Val\`{e}ncia'' team. It uses a U-Net convolutional neural network whose main hyperparameters have been optimized to obtain the best AEC performance, but limiting the size of the network in order to meet the latency restriction of $40$\,ms imposed by the challenge's rules. For the network training and testing, we have used a synthetic dataset (S-U-Net) and a synthetic dataset together with the single-talk set of a real dataset (SR-U-Net), both datasets were provided by the challenge organizer. The S-U-Net model presented better results for double-talk scenarios and their results were submitted to the challenge, obtaining an overall MOS of $3.57$ and ranking 12th among 17 teams.

\section*{Acknowledgment}
This work has been partially supported by grants RTI2018-098085-B-C41 (MCIU/AEI/FEDER) and PROMETEO/2019/109. M.F.~Dolz work was also supported by the Plan GenT project CDEIGENT/2018/014 of the \emph{Generalitat Valenciana} and J.~Silva-Rodríguez work was also supported by the Spanish Government under FPI Grant [PRE2018-083443].






\bibliographystyle{IEEEtran}
\bibliography{biblio.bib}

\end{document}